\documentclass[%
 reprint,
 amsmath,amssymb,
 aps,
]{revtex4-1}

\usepackage{graphicx}
\usepackage{dcolumn}
\usepackage{bm}
\usepackage{hyperref}

\usepackage{braket}
\usepackage{mathtools}
\usepackage{stmaryrd}
\usepackage{mathrsfs}
\usepackage{float}
\usepackage{natbib}
\usepackage{tensor}

\begin{document}

\preprint{APS/123-QED}

\title{Exotic Behaviors of Space-time in Einstein-Maxwell Theory}

\author{Erik W. Lentz}
\email{erik.lentz@uni-goettingen.de}
 \affiliation{Institut f\"ur Astrophysik, Georg-August Universit\"at G\"ottingen, }

\date{\today}

 \begin{abstract}

This paper examines the structure of electric fields and space-times created by extended finite distributions of irrotational static and spherically-symmetric charge. The resulting electric fields are found to source features in space-time commonly associated with the presence of fields with locally positive and negative mass densities, with the sign of the mass corresponding to the sign of the electric field gradient. The conditions detailing these behaviors are discussed. The effects of these structures on trajectories of time-like geodesics and the volume form are also presented.

%

\end{abstract}

\maketitle


\section{Introduction}

Classical electric and magnetic fields source stress-energy in space-time, which in turn affects the geometry of dynamical space-time theories such as Einstein's general relativity \citep{Weinberg1972,Wald1984}. The geometry of space-time is also known to induce structure in the electric and magnetic fields, including responses analogous to polarization, magnetization, and in general a suite of magneto-electric effects \cite{PhysRev.97.511,Landau1975, 2019arXiv190311805C, 2019Univ....5...88G,Zelmanov2006,2019Univ....5...88G,2019arXiv190311805C}. Solutions to Einstein-Maxwell theory were found almost immediately following the announcement of general relativity. These solutions have broadened greatly in scope over the last century, though the Maxwell contribution most often plays a sub-dominant role to other massive fields \citep{1916AnP...355..106R,1917AnP...359..117W,1918KNAB...20.1238N,doi:10.1098/rspa.1921.0028}, or is made into an incoherent radiation fluid \citep{Weinberg1972,Wald1984}. 

General relativistic systems with stress-energy dominated by coherent classical electromagnetism are rarely studied. Though their space-times have vanishing curvature scalar due to the trace-less nature of the scale-invariant Maxwell stress-energy tensor, their geometries are far from trivial. The light-like electromagnetic stress-energy has been demonstrated to produce geometries with features similar to those produced by fields with positive and negative mass densities, such as the Reissner-Norstr\"om metric \cite{1916AnP...355..106R,1917AnP...359..117W,1918KNAB...20.1238N,doi:10.1098/rspa.1921.0028,1983PhLA...99..419Q}, where the repulsive quality of a spherical electric field sourced by a point charge can be likened to a singular source of negative mass surrounded by a positive mass density with extent comparable to the radial electric field. Despite this curious finding, the solutions of pure Einstein-Maxwell theory remain largely unexplored.

I present in this paper example solutions of Einstein-Maxwell theory with the purpose of examining the exotic mass-like response of space-time to electric fields. The space-time metric and Maxwell fields studied here are derived from prescribed static, spherically-symmetric, and irrotational charge distributions of large extent, which generalize the solutions of the charge-dominated Reinser-Nordstr\"om system. Non-singularity of the charge distributions are expected to reveal structures not visible in the charged back hole solution. The remainder of the paper will be organized as follows: Section~\ref{theory} states the Einstein-Maxwell equations and presents the forms satisfying the prescribed symmetries; Section~\ref{examples} presents solutions of the derived equations for multiple sources of charge, in configurations with charge with a single sign and of both signs; Section~\ref{discussion} discusses in what ways the electric field stress-energy can emulate matter with either positive or negative mass and the conditions for each on the electric field; and Section~\ref{summary} summarizes the results and outlines potential for further study.

\section{Theory }
\label{theory}

The field equations of Einstein-Maxwell theory can be derived from a single action over the smooth metric-compatible space-time manifold $(\mathcal{M},g)$ with Lagrangian density $\mathcal{L}_{EM} = R/2 \kappa - F^2/4 \mu_0$, where $R$ is the Ricci curvature tensor, $\kappa$ is the Einstein constant, $F^2$ is the modulus squared of the electromagnetic field strength, and $\mu_0$ is the vacuum magnetic permeability. The Euler-Lagrange equations with respect to variations in the metric provide the Einstein equation, which has components
\begin{equation}
    G^{\mu \nu} = \kappa T^{\mu \nu},
\end{equation}
where $G^{\mu \nu} = R^{\mu \nu} - g^{\mu \nu} R/2$ are the components of the Einstein tensor, $R^{\mu \nu}$ are the components of the Ricci tensor, $R$ is the Ricci scalar, and the stress energy tensor components are derived to be
\begin{equation}
    T^{\mu \nu} = F^{\mu \alpha} F^{\nu} _{\alpha} - \frac{1}{4} g_{\mu \nu} F_{\alpha \beta} F^{\alpha \beta}.
\end{equation}
The Maxwell stress energy naturally satisfies numerous energy conditions including the dominant, weak, and null types \cite{Wald1984}. The remainder of this paper will be presented in units where $c=\kappa=1$.

The covariant Maxwell equations can be found from a combination of variational techniques and a Bianchi identity. The variational Maxwell equations are given by
\begin{equation}
    \nabla_{\mu} F^{\mu \beta} = -J^{\beta},
\end{equation}
where $\nabla$ is a covariant derivative and $J^{\beta}$ are the components of the electric current density. The current density has been inserted by hand to minimize its contributions to the stress-energy, and will serve as a prescribed quantity without dynamics. A Bianchi identity appears in Maxwell theory as the field strength is an exact differential form, causing the field strength's exterior derivative vanish. In coordinates this looks like
\begin{equation}
    \nabla_{\alpha} F_{\beta \gamma} + \nabla_{\gamma} F_{\alpha \beta} + \nabla_{\beta} F_{\gamma \alpha} = 0.
\end{equation}

The solutions presented here are limited to those space-times that are irrotational and exhibit both time translation symmetry and spherical symmetry. Such restrictions reduce the line element in polar spherical coordinates to
\begin{equation}
    ds^2 = h(r) dt^2 - \gamma_{rr}(r) dr^2 - r^2 g_{\omega},
\end{equation}
where $g_{\omega}$  gives the metric of the unit two-sphere. Only the radial electric field survives among the electromagnetic degrees of freedom. The relevant geometric equations can be picked from the time-time component of the Einstein equation and the null curvature constraint
\begin{align}
    &R^{00} = T^{00} = \frac{1}{2} E^r(r)^2 \gamma_{rr}, \\
    &R = 0.
\end{align}
The only relevant Maxwell equation here is Gauss' law, which for the radial electric field becomes
\begin{equation}
    \frac{1}{\sqrt{\gamma_{rr}}} \partial_r \left( \sqrt{h \gamma_{rr} } E^r(r) \right) =  \rho.
\end{equation}
Charge distributions here are made to occupy non-zero volume to avoid the singularity of the Reissner-Nordstr\"om solution.

\section{Results}
\label{examples}

\subsection{Homogeneous Sphere}

\begin{figure}
\begin{center}
\includegraphics[width=\columnwidth]{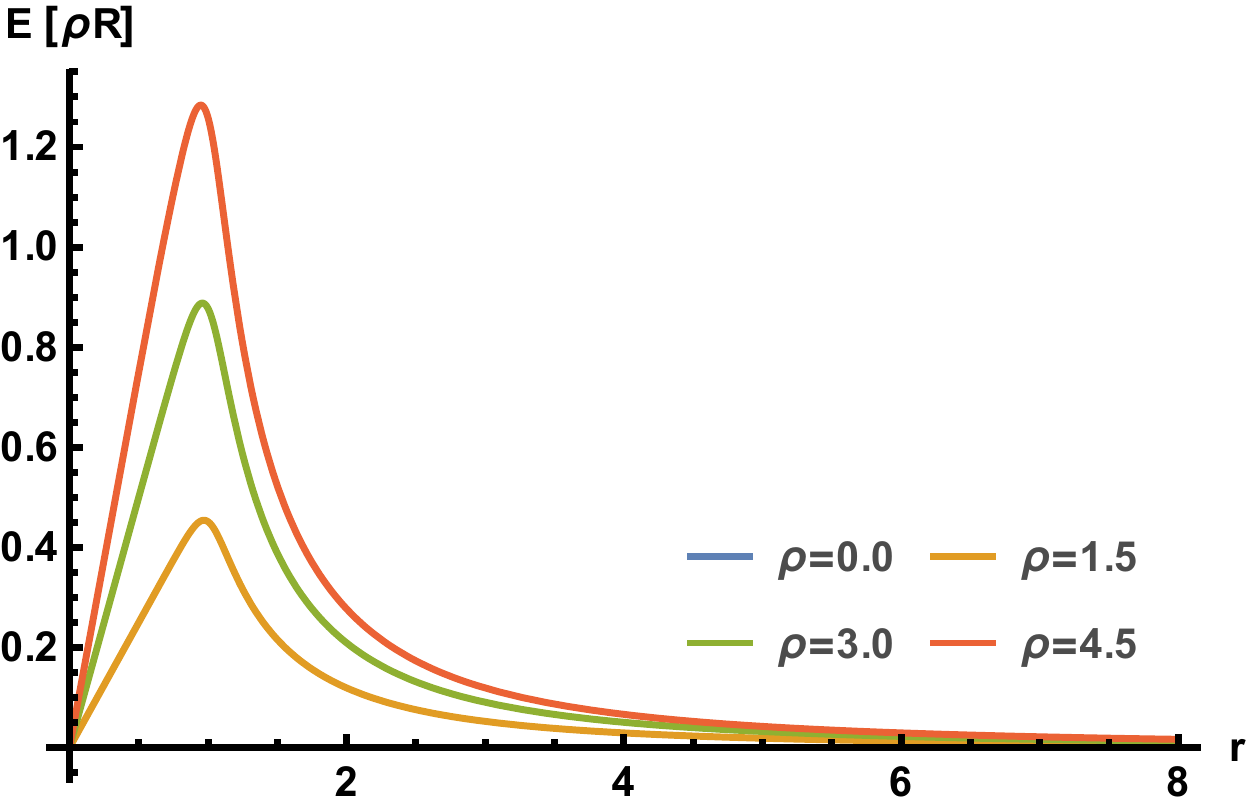}
\caption{Radial electric field over radius of the homogeneously-charged sphere.}
\label{fig:DEP}
\end{center}
\end{figure}

The first charge distribution is a finite sphere of nearly homogeneous charge density
\begin{equation}
\rho_H (r) =  \frac{\rho}{2} \left(1 + \tanh \left( (R-r)/\sigma \right) \right)
\end{equation}
where the step-function edge is smoothed here by a hyperbolic tangent of width $\sigma/R=0.1$. Recall that the charge densities here have no dynamics of their own and are held as a prescribed source for the electric field. The behavior of the resulting electric field approximately follows the flat space-time solution, building in a near-linear fashion within the sphere, and decaying at $\sim 1/r^2$ far outside the sphere's edge ($r-R>>\sigma$), Fig.~\ref{fig:DEP}. The form of the field remains largely the same as the sphere's charge increases, with strong-field changes coming in the form of a diminished growth of the field within the sphere, a faster decay near the sphere's outer edge, and a general diminishing of the field's amplitude relative to the linear response of the flat space-time case.

\begin{figure*}
\begin{center}
\includegraphics[width=\textwidth]{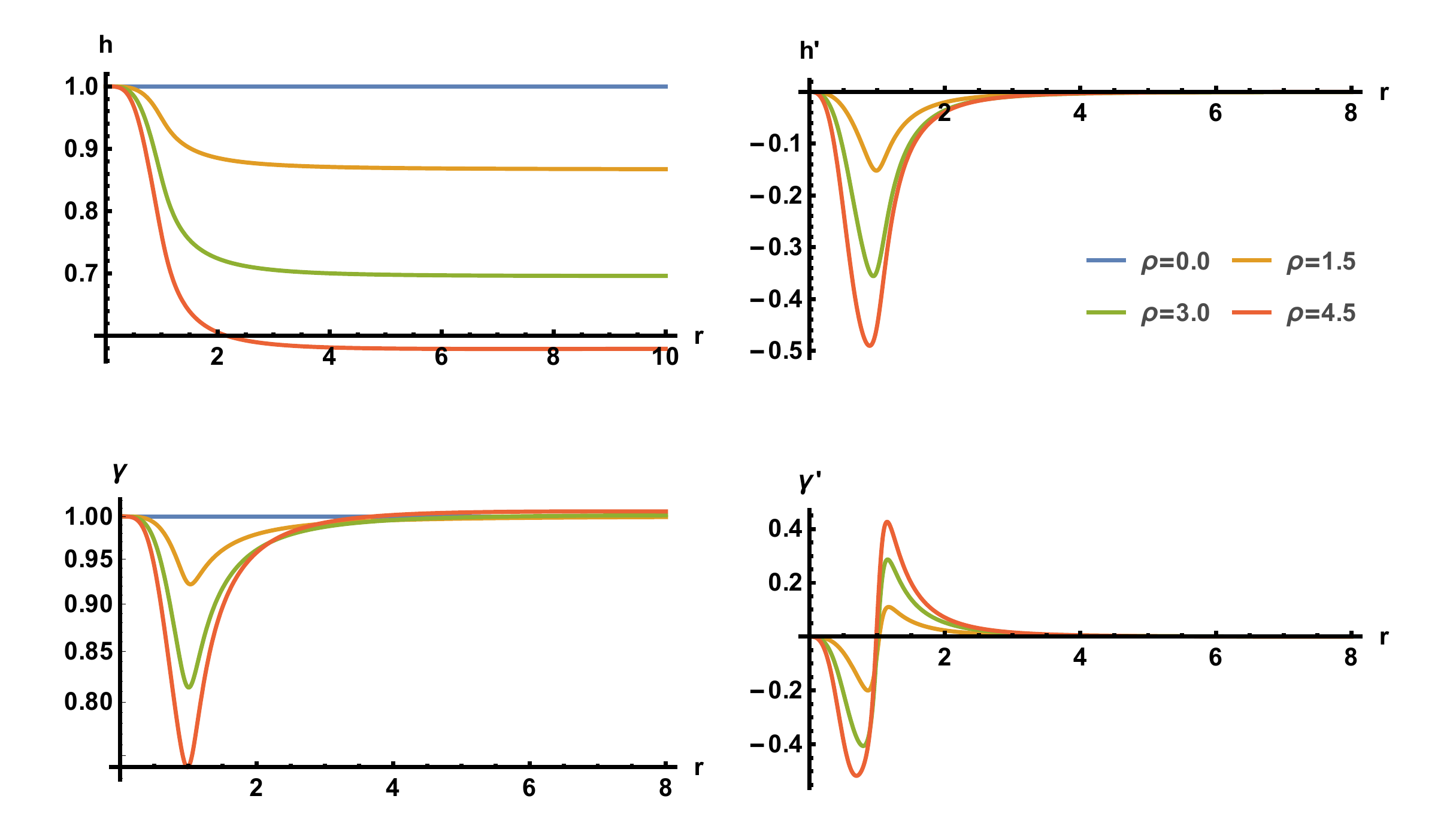}
\caption{Metric components $h$ (upper left) and $\gamma_{rr}$ (lower left) and their derivatives with respect to radius (upper right and lower right respectively) for the homogeneously-charged sphere. Both metric components maintain a flat space-time at the sphere's center as well as asymptotically far from the outer shell. Note that the measures in and out of the sphere deviate as the charge grows. The components of the metric have been normalized to unity at $r=0$ to better highlight the differences in structure within the charge distribution. A simple re-scaling of the components can be performed to regain the standard form of the asymptotic Minkowski metric. }
\label{fig:gsphererho}
\end{center}
\end{figure*}

The enclosed energy of the electric field depresses the component $h$ as one increases in radius, Fig.~\ref{fig:gsphererho}. The geometries of the homogeneous sphere charges take the shape of the mass-less Reissner-Nordstr\"om metric far from the edge of the charged sphere \citep{doi:10.1098/rspa.1921.0028}, asymptoting toward the infinite-radius value with as an inverse square $h(r) - h(\infty) \propto Q^2/r^2 $. The $h$ component's behavior is matched in reciprocal in this region by the radial metric component $\gamma_{rr}$. Both components quickly converge to their infinite range values, localizing the curvature of the space-time. The metric changes its behavior closer to the sphere and within its interior. The now radially-increasing electric field induces opposite behavior in $\gamma_{rr}$, which changes sign in slope and now is decreasing with radius. The component $h$ maintains its original slope, inflecting towards flatness at the origin. The metric components here have diverged from their reciprocal relationship observed far outside the sphere. The center of the metric is flat, restricting the region of non-trivial curvature to an annulus in space-time, though the measures of time and radius in these coordinates are mismatched from their $r\to\infty$ counterparts. 

The time-like geodesics of these space-times are generally repelled from the sphere, save for the single unstable stationary point at the origin. The effective radial gravitational force for time-like geodesics with a coordinate-stationary initial condition is given by
\begin{equation}
    F^r = -\Gamma^r_{tt} = -\frac{\partial_r h}{2 \gamma_{rr}},
\end{equation}
and shows a growing repulsive force in the sphere's interior, and a force shrinking faster than $r^{-2}$ outside the sphere.

\subsection{Sphere with Charge Sign-Flip}

The second distribution shape presented here oscillates the sign of the charge density within the sphere. Here the charge density is modified by a half cosine function
\begin{equation}
\rho_O (r) =  \rho_H(r) \cos \left( \pi r \right).
\end{equation}
The sphere is not neutral by design. The charges are configured such that the electric field may flip direction within the sphere's boundary. The flip does not change the sign of the energy density of the electric field, which is by construction positive definite.

\begin{figure*}
\begin{center}
\includegraphics[width=\textwidth]{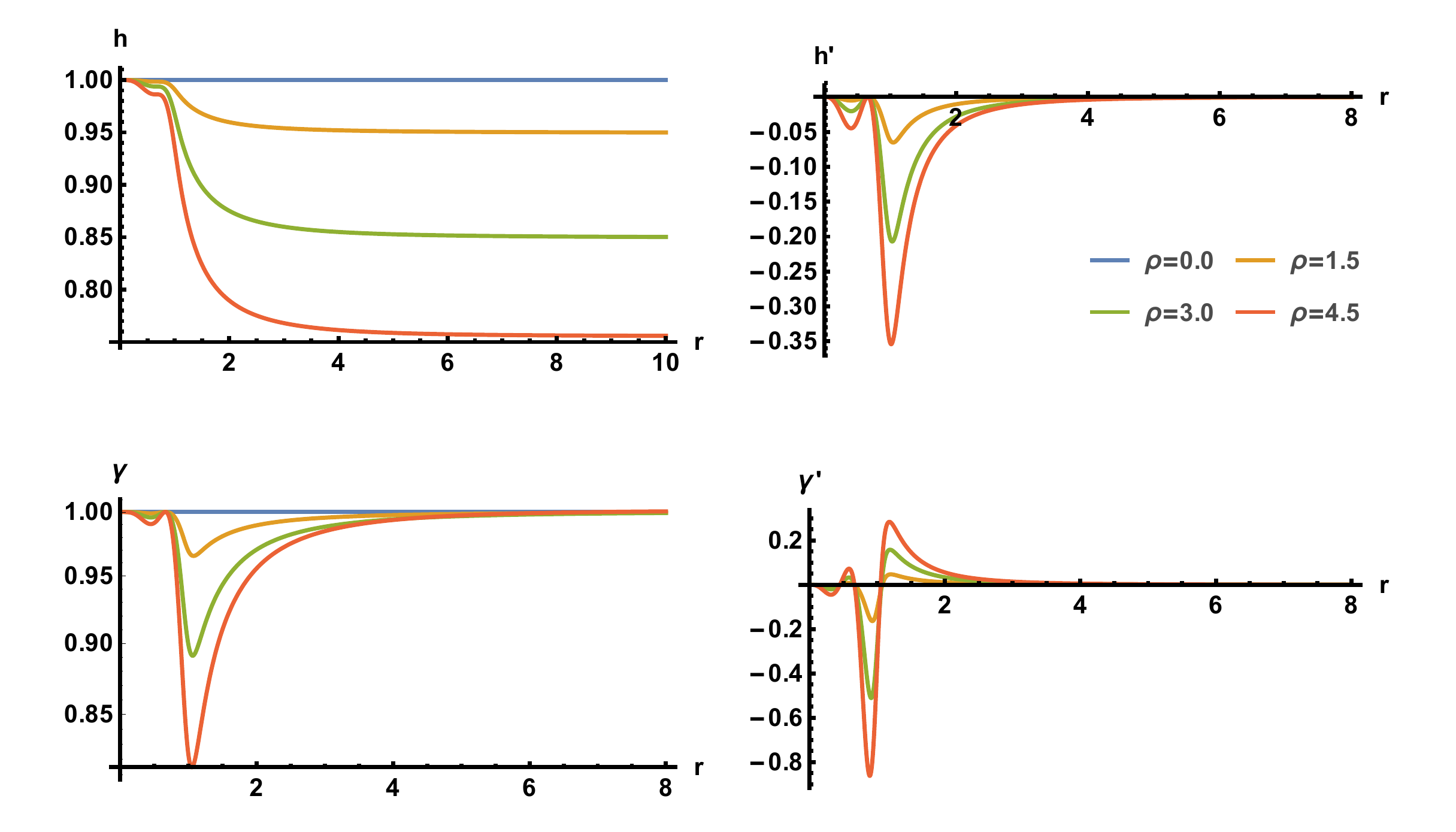}
\caption{Metric components $h$ (upper left) and $\gamma_{rr}$ (lower left) and their derivatives with respect to radius (upper right and lower right respectively) for the sphere with sign-flipping charge density. Metric components still maintain a flat space-time at the sphere's center and  asymptotically far from the outer shell, but with more complex interior behavior than the homogeneous sphere. The components of the metric have been normalized to unity at $r=0$ to better highlight the differences in structure within the charge distribution. A simple re-scaling of the components can be performed to regain the standard form of the asymptotic Minkowski metric. }
\label{fig:gsphererhoosc}
\end{center}
\end{figure*}

The metric far from the sphere looks similar to the homogeneous sphere, seeing only the magnitude of the sphere's total charge, Fig.~\ref{fig:gsphererhoosc}. The shape of the metric components begin to differ from the homogeneous sphere within the sphere's boundary. The oscillating charge distribution produces additional structures in both $h$ and $\gamma_{rr}$. Specifically, the $h$ component forms a small shelf within the sphere, extending between the charge-density zero-point at $r=0.5$ and $r \gtrsim 0.7$. The center of the sphere behaves similarly to the homogeneous example due to their matching central charge densities. The plot of $h'$ shows the shelf has a single flat point. The $\gamma_{rr}$ component shows a similar structure to $h'$ within the sphere, producing a number of oscillatory features in the slope before relaxing smoothly to a flat central state. The non-center zero point in $h'$ specifically matches a feature in $\gamma_{rr}$ where the component smoothly, and momentarily, resumes the vacuum state. 

\begin{figure}
\begin{center}
\includegraphics[width=\columnwidth]{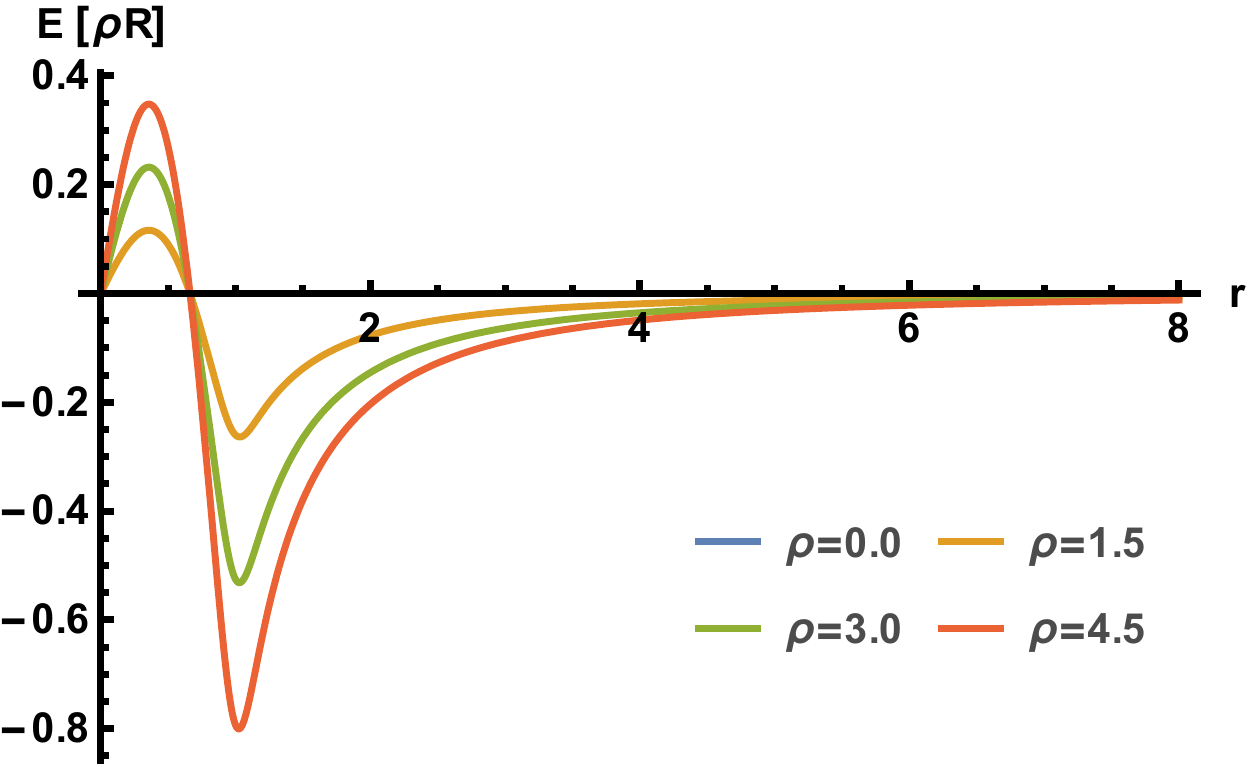}
\caption{Radial electric field over radius of the sphere with sign-flipping charge density. Note the stationary nature of the interior zero field point. }
\label{fig:DEPosc}
\end{center}
\end{figure}

The electric field produces a familiar structure, Fig.~\ref{fig:DEPosc}. The radius at which the electric field vanishes within the sphere $r_{root}$ matches the radius at which the enclosed charge also vanishes, as expected by the integral form of Gauss' law. This root also coincides with the flat point on the $h$ shelf. Neutral spheres with similar but smaller charge density to $\rho_0$ show similar structure. 

\begin{figure}
\begin{center}
\includegraphics[width=\columnwidth]{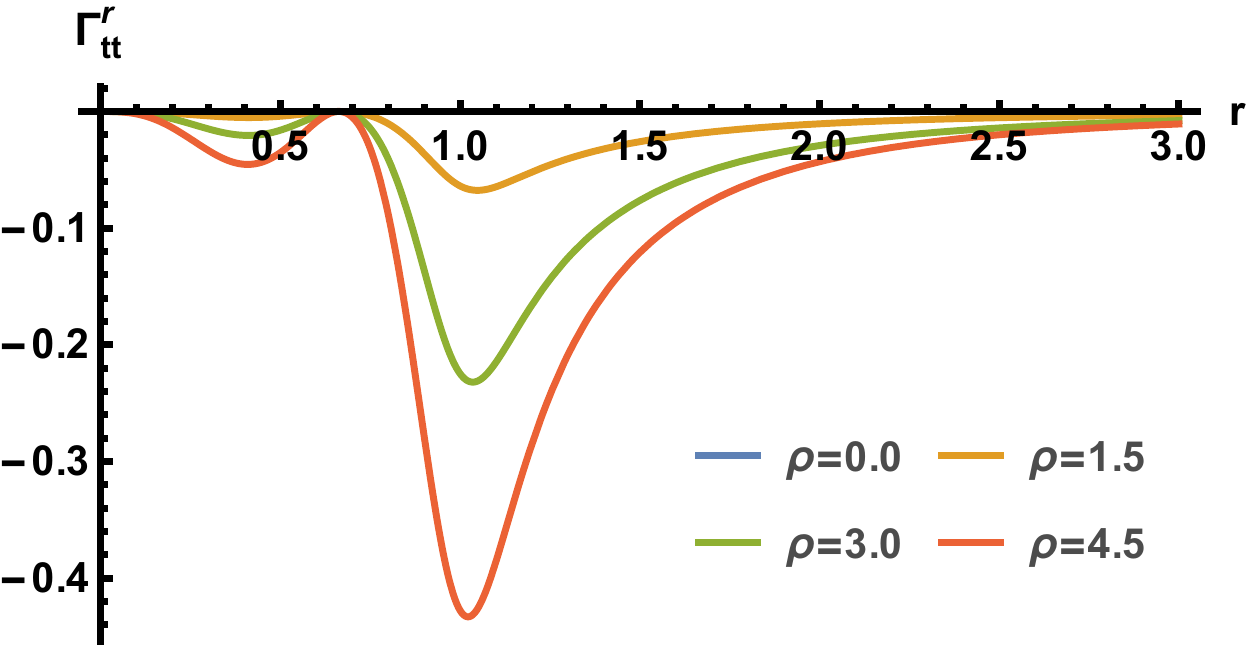}
\caption{Time-radius-radius Christoffel symbol over radius for the sphere with sign-flipping charge density. This symbol evaluates to the negative of the effective force of gravity on a space coordinate stationary world line of a time-like geodesic. }
\label{fig:force_osc}
\end{center}
\end{figure}

The homogeneous sphere's has the characteristic that time-like geodesics have a stationary point at infinite radius and an unstable point at the origin. The charge sign-flipping sphere has an additional stationary point within the sphere, occurring at $r_{root}$, Fig.~\ref{fig:force_osc}. The new point is quasi-stable, with a restoring force for negative displacements but a runaway force towards infinity for positive displacements. Neither a sphere of charge with a single sign nor a mass density of a single sign can produce this second stationary point.

\begin{figure*}
\begin{center}
\includegraphics[width=\textwidth]{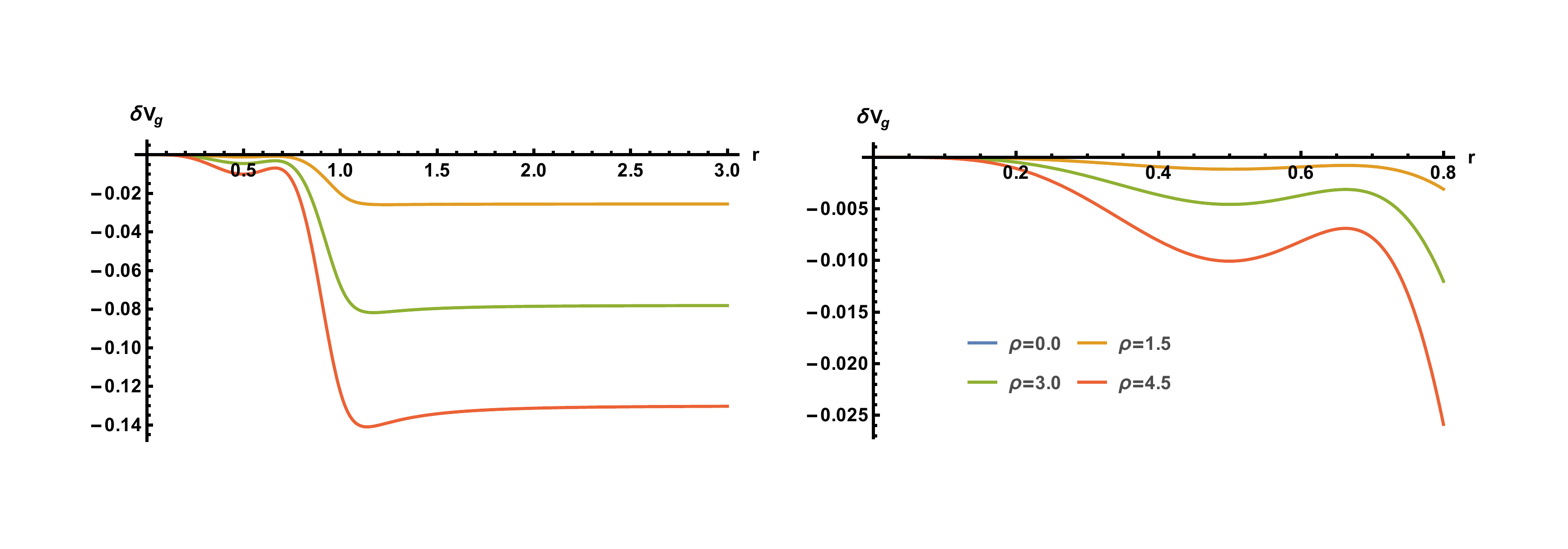}
\caption{Difference in volume measure $\delta V_g$ over radius for the sphere with sing-flipping charge density. The left sub-figure displays $\delta V_g$ out to three times the sphere radius, while the right figure is a zoom in of the sphere's interior. }
\label{fig:deltaV_osc}
\end{center}
\end{figure*}

The shelf region also creates an unusual feature in the volume measure, Fig.~\ref{fig:deltaV_osc}. The fluctuations in the volume scalar relative to the Minkowski metric display local expansions where the electric field is decreasing. Here we use the form modulated by the sphere surface area in spherical coordinates
\begin{equation}
    \delta V_g = \frac{ \sqrt{|g|} - \sqrt{|g_{Mink}|}}{r^2 \sqrt{|g_{\Omega}|}},
\end{equation}
which increases the visibility of the feature, though it exists as a maximum in the full scalar as well. These expansions in the volume occur in two places: in the region where negative electric charge is accumulating to neutralize the enclosed charge, maximizing at $r_{root}$, and outside the sphere where the field decays to vacuum. A similar expansion feature is also seen outside of the homogeneous sphere.

\section{Discussion}
\label{discussion}

The exotic effect of the electric field stress-energy on geometry that has been known since the time of Reissner and Nordstr\"om. The repulsive nature of the electric stress-energy emulates that of a field with negative mass density. One can make this link more evident by constructing an enclosed-mass analogue for the electric contribution to space-time based on the Reissner-Norstr\"om solution
\begin{equation}
    M^{eff}_<(r) \sim -\frac{Q_<(r)^2}{r}
\end{equation}
where $Q_<$ is the enclosed charge. The effective enclosed mass figure over the interior of the homogeneous sphere is monotonically decreasing, as would be expected by a distribution of negative mass density. Outside the sphere, however, the enclosed mass figure's slope changes sign, like the sign of the electric field magnitude's slope. The effective mass figure begins to asymptotically approach zero, acting as if there is a source of positive mass density outside the sphere. This is due to the extended nature of the electric field. One may intuit from this perspective that a radially increasing (in magnitude) electric field appears like a negative mass density, while a decreasing electric field magnitude emulates a positive mass density. This rule is strengthened by the findings of the sign-changing charged sphere.


The sign-flipping sphere allows one to experience both the positive and negative mass features of an electric-field-dominated space-time in a finite volume. One can even restrict the curvature of space-time to a compact region by producing a sphere of net zero charge. It is essential that electric charge has both positive and negative signs for this to occur in a finite volume. The point of vanishing electric field  $r_{root}$ and enclosed charge also leaves its mark on the geodesic force and the volume measure. The second finite-radius stationary point is a stand-out feature of the gauge-field-dominated space-time. Such a non-central stationary point cannot occur with a stress energy from a massive field as the enclosed energy $M^{eff}_<(r)$ must monotonically grow as mass charge has only been realized with a positive sign. This implies that one would need sources of positive and negative mass density to reproduce such an outer stationary point. Classical electromagnetism can serve as both in this way.


The volume scalar also experiences a interruptions in its monotonic behavior for the observed spheres. For the sign-flipping charged sphere, this change occurs between the polarity-flipping point of the charge density at $r=0.5$ and continues to the enclosed charge zero point at $r=r_{root}$. This annulus is where one would expect a band of massive material to lie. Another uptick in the volume measure occurs outside the sphere, being most visible among the stronger fields. This corresponds to a similar positive mass contribution that takes place on the tail of the electric field. Overall, space-time appears to respond in several ways to growing electric fields as if the region contains negative mass density, and to shrinking electric fields as if the region had positive mass density. Both dilation and contraction of the volume form is not expected from a static massive field.

\section{Summary}
\label{summary}

This paper studies the properties of static spherically-symmetric irrotational electric charge distributions and the resulting electric fields and space-times via Einstein-Maxwell theory. The charge distribution types studied there include a homogeneous sphere and a sphere with charge density modulated by a half-cosine function. Solutions show that space-time and the electric field follow the mass-less Reissner-Nordstr\"om solution far outside the charged volume. The exterior solutions are matched according the their net charge. The electric field and metric inside the charge distribution smoothly connect to a flat state at the origin, but with additional structure that cannot be seen in the singular Reissner-Nordstr\"om case. The interior structures include curvatures that emulate some of the responses expected of a space-time subject to stress-energy with mass, where the mass density is in one region positive and in another negative. The properties present themselves in the behavior of time-like geodesics and the volume scalar. The conditions for this behavior appear to nearly correspond to the sign of the radial gradient of the electric field, with positive gradients corresponding to negative mass density, and negative gradients to positive mass density. The negative mass-like behavior can also be seen outside the sphere, as the electric field relaxes to zero. These findings match with those of the Reissner-Nordstr\"om case if the extent of the charge distribution is taken to zero. The bi-polar nature of electric charge plays an important role in inducing these structures for finite volumes.

The ability of static space-times with electromagnetism to emulate properties of space-times containing positive and negative mass density is so far unique among observed macroscopic fields. These results are taken to imply that there is great potential in Einstein-Maxwell theory to discover varieties of space-times that so far have only been theorized using forms of matter not yet realized. This potential will be studied in greater detail in future work, first by reducing the symmetries imposed on the configurations, as well as incorporating dynamics of the charge distribution.

\section{Acknowledgements}

I would like to thank Katy Clough for productive discussions regarding this topic.

\bibliography{main.bib}

\end{document}